\documentclass[twocolumn,superscriptaddress,prb,showpacs]{revtex4}
\usepackage{amsbsy,amssymb,amsmath,bm,graphicx,subfigure}

\begin{document}
\title{Spin echo without an external permanent magnetic field}
\author{Joakim Bergli}
\email{jbergli@fys.uio.no}
\affiliation{Physics Department,
University of Oslo, PO Box 1048 Blindern, 0316 Oslo, Norway}
\author{Leonid Glazman}
\email{glazman@umn.edu}
\affiliation{Theoretical Physics Institute,
University of Minnesota,
Minneapolis, MN 55455
}

\begin{abstract}
The spin echo techniques aim at the elimination of the effect of a 
random magnetic field on the spin evolution. These techniques conventionally
utlize the application of a permanent field which is much stronger than 
the random one. The strong field, however, may also modify the 
magnetic response of the medium containing the spins, thus altering their
``natural'' dynamics. We suggest an iterative scheme for generating a 
sequence of pulses which create an echo without an external permanent field.
The approximation to the ideal echo improves with the 
sequence length. 
\end{abstract}

\pacs{76.60.Lz}

\maketitle

\section{Introduction}

The use of  echoes is a standard technique in spin resonance experiments
\cite{Cowan}, where a large number of echo pulse sequences are
used. Echo sequences typically consist of a number of high-frequency
 pluses that induce controlled rotations on the precessing
spin. The objective is to remove or reduce dephasing due to
inhomogeneities in the external magnetic field which give a variation
of the precession rate across the sample (known as inhomogeneous
broadening). Since any two-level quantum system can be mapped on a
spin-$\frac{1}{2}$ in a magnetic field, similar operations can be
performed on any such system, including a qubit. For example, there are recent
experiments on superconducting qubits \cite{Nakamura2002,Vion2003}
(were it is called charge echo, since the physical degree of freedom is 
 two charge
states of a superconducting grain) and electron spins in quantum dots
\cite{Petta2005}. Common to the existing echo techniques
 is the restriction that in
order to apply the proper sequence of control rotations one must know
the direction of the magnetic field. In other words, the sequences are
able to remove dephasing due to inhomogeneities in the field
amplitude, but not the field direction (the randomness in the direction of  
the magnetic field is indeed negligible if a strong external field is 
applied). Starting with the experiments
\cite{Kikkawa1998} that showed long spin coherence times of electrons
in bulk GaAs there has been interest in using electrons confined in
quantum dots as qubits. It appears that the limiting decoherence
factor in these systems is the hyperfine interaction between the
electron and nuclear spins \cite{Khaetskii}. As long as no special
preparation is made, the nuclear spin system will give rise to an
effective random magnetic field seen by the electron. This field
will be random both in strength and direction, and in the absence of
an external strong permanent magnetic field the usual echo sequences
can not be applied. Because the response of the nuclear spin system is
much slower than the precession of the electron spin we can in a
certain approximation assume the nuclear field to be constant.
Motivated by this we will in this paper investigate the problem of
finding an echo sequence that is applicable in the case of a constant
effective field with an arbitrary and unknown magnitude and direction.
The goal of this sequence is to return the spin to its starting
position at the end of the sequence, independently of the magnetic
field.

\section{Formulation of the problem}

Consider a spin-$\frac{1}{2}$ particle precessing in a constant magnetic
field. Let ${\bf n}$ be the unit vector along the precession axis and 
$\omega$ be the precession angular frequency. The rotation of the spin
state during  time $\tau$ is then given by the unitary operator

\begin{equation}\label{U}
 U = \cos\chi I + i\sin\chi n_i\sigma_i,
\end{equation}
where  $I$ and $\sigma_i$ are the identity and Pauli matrices, and 
$\chi = \omega\tau/2$.
The $\pi$ rotations about the coordinate axes are denoted 
$X=i\sigma_x$, $Y=i\sigma_y$, and $Z=i\sigma_z$.

The simplest form of a spin echo is for the situation where the spin 
is precessing in a field with known direction, say along the $z$-axis, 
but with unknown magnitude. Then $n_x=n_y=0$ and $n_z=1$, while $\chi$
is arbitrary. The usual echo sequence then consists in waiting for the 
time $\tau$, applying an $X$ rotation, waiting time $\tau$ and applying
a final $X$ rotation. The success of the procedure is expressed by the 
fact that $XUXU=-I$ is the identity (up to a global sign, 
which is unimportant),
independently of $U$. By symmetry, the same is true if $X$ is replaced by 
a $\pi$ rotation around any axis in the $xy$-plane (perpendicular 
to the precession axis). In particular we have $YUYU=-I$. 

We want to extend this to the case were ${\bf n}$ is not known. That is, 
we want to find a sequence of control rotations $A,B,C,\ldots,F$ such that 

\begin{equation}\label{F}
FU\cdots CUBUAU = I
\end{equation}
for any $U$. We make the following assumptions: i) The control 
pulses are effectively instantaneous, meaning they can be performed in a time
much less than the precession period. ii) The external magnetic field is 
unknown but constant, so that the precession operator $U$ does not 
change in time. Note that we have written Eq. (\ref{F}) as if the  
time $\tau$ between the pulses is fixed. Different time intervals 
between the pulses are achieved by choosing some of the $A,B,C,\ldots,F$ 
to be the identity. This is sufficient if all intervals are integer multiples
of a smallest unit. Intervals with irrational ratios will require a more 
general form than Eq. (\ref{F}). In this paper we only use equal intervals.

\section{Iterated mappings}

As was explained above, if we know the direction ${\bf n}$ of the external 
field we can create an echo by applying two $\pi$ pulses around any 
axis perpendicular to ${\bf n}$. The sequence $XUXU$ can 
be used if $n_x=0$, and similarly we may use $YUYU$ if we know that $n_y=0$. 
Consider the longer sequence $XUYUXUYU$ for which, using the general 
$U$ of Eq. (\ref{U}), we get 

\begin{align}\label{it}
 &XUYUXUYU \\ &= (8n_x^2n_y^2\sin^4\chi-1)I  
   -8in_xn_y^2\sin^3\chi\cos\chi\sigma_x \nonumber 
  \\&-8in_xn_y^2\sin^4\chi\sigma_y
    -4in_xn_y\sin^2\chi(1-2n_y^2\sin^2\chi)\sigma_z. \nonumber
\end{align}
We see that $XUYUXUYU=-I$ for either $n_x=0$ or $n_y=0$. This 
may seem like a small gain, but this sequence is the key to the full solution. 
Let us construct the mapping

\begin{equation}\label{T}
  U\rightarrow T(U)= XUYUXUYU.
\end{equation}
The idea is that $T(U)$, being composed of the arbitrary $U$ and the fixed
control rotations, will be ``less arbitrary'' than the original $U$. 
Iterating this mapping we then construct the set 
$U^{(1)} = T(U)$, $U^{(2)} = T(U^{(1)})$, ... of pulse sequences that 
will be better and better approximations to the identity. Since 
$T(U^{(n)})$ contains $U^{(n)}$ four times the time needed for the sequence
$U^{(n)}$ is $4^n\tau$. That is, the length of the sequence grows 
exponentially in $n$. This is of course unfortunate as the sequences
quickly will become impractically long. However we will see below that 
for a large portion of the space of parameters determining $U$, 
a few iterations $n$ are sufficient to reach a good approximation to the 
identity. 
The domain of ``bad''
 parameters shrinks exponentially with the increase of $n$.

To illustrate this we represent a rotation by the polar angle $\theta$
and azimuth angle $\phi$ of the rotation axis (so that $n_x=\sin\theta\cos\phi$,
$n_x=\sin\theta\sin\phi$ and $n_z=\cos\theta$) and the rotation angle $\chi$.
\begin{figure}
  \begin{center}
 \subfigure[$U$]{\includegraphics[width=4cm]{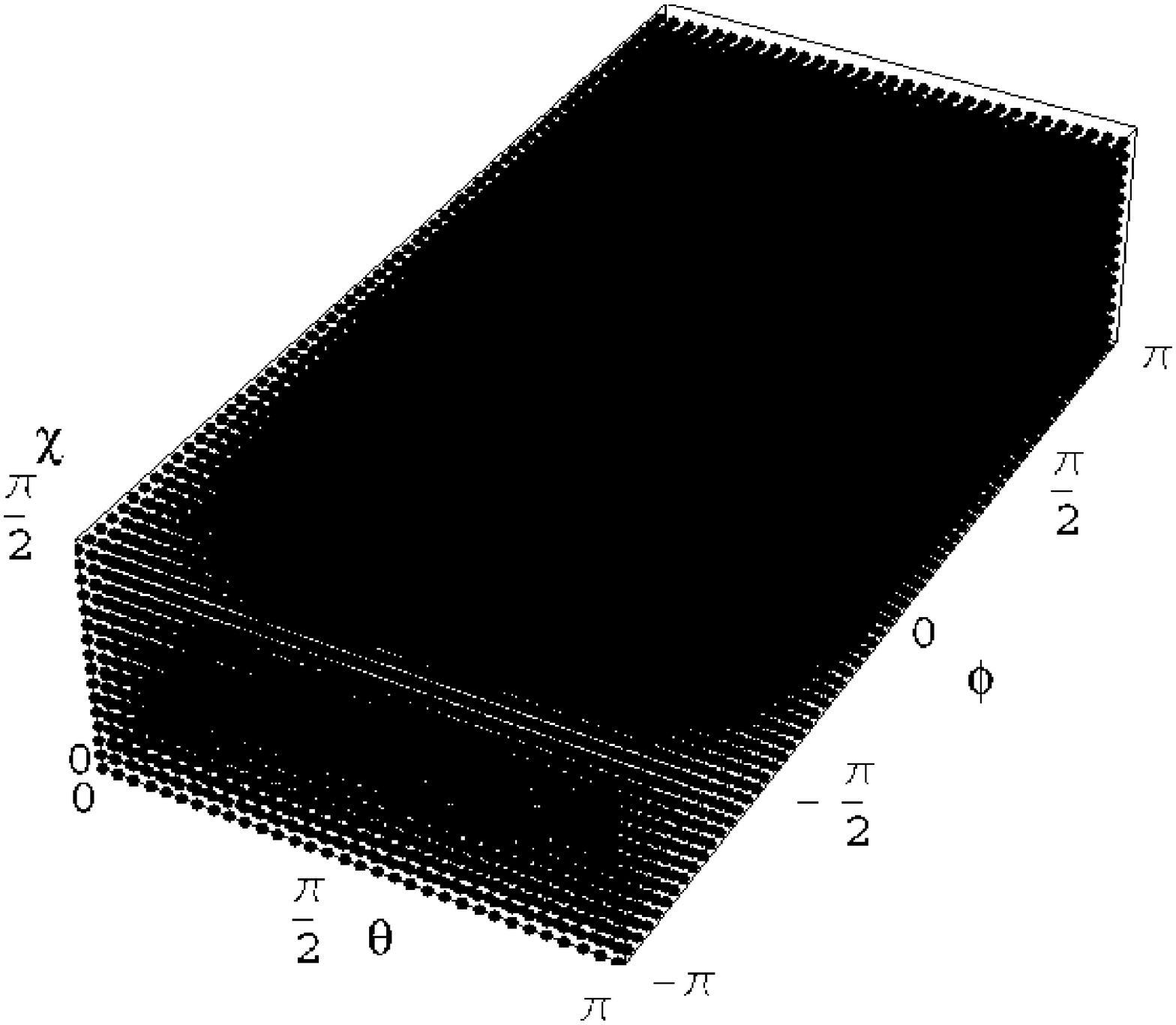}}
 \subfigure[$U^{(1)}$]{\includegraphics[width=4cm]{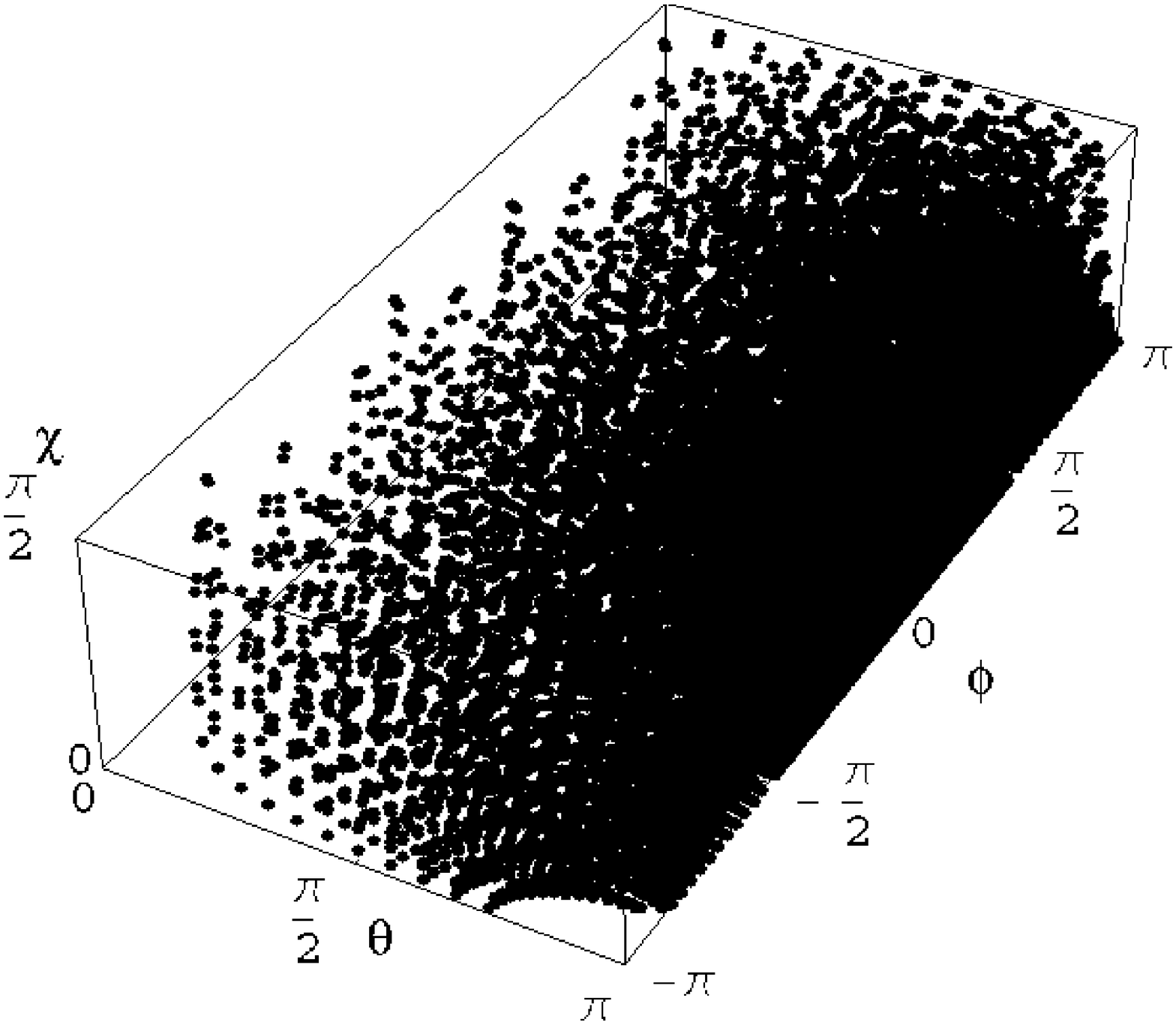}}\\
 \subfigure[$U^{(2)}$]{\includegraphics[width=4cm]{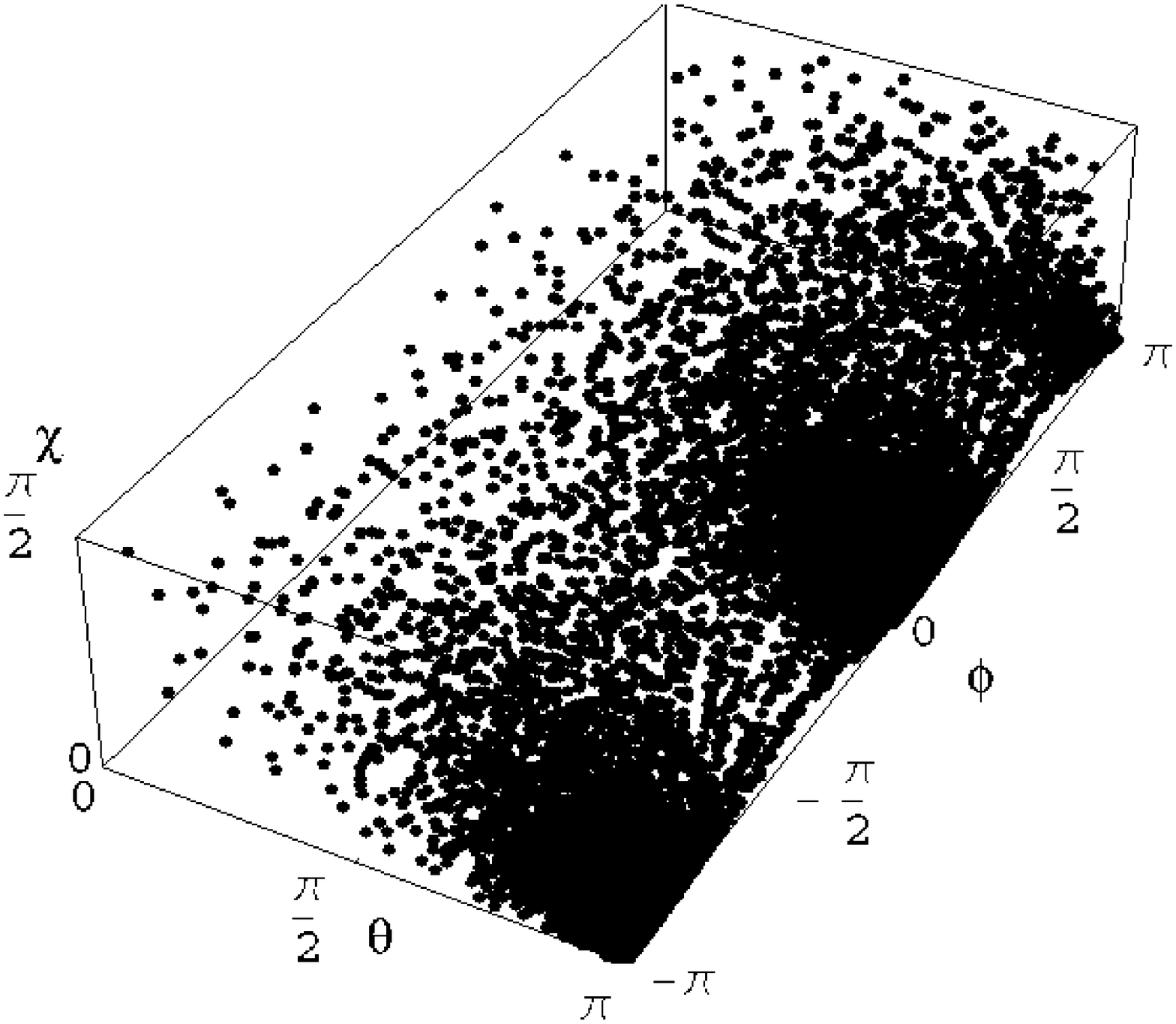}}
 \subfigure[$U^{(3)}$]{\includegraphics[width=4cm]{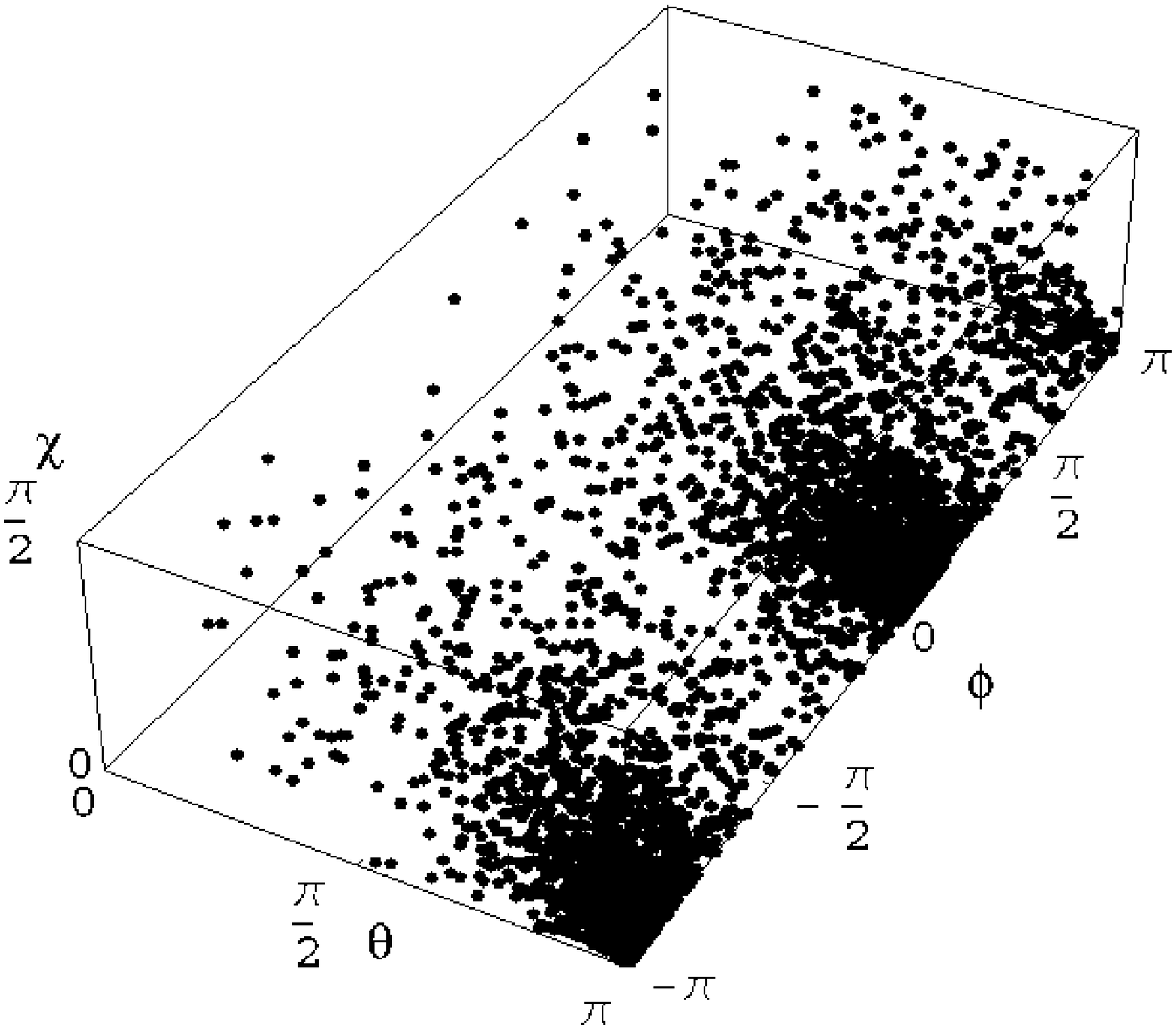}}\\
 \subfigure[$U^{(4)}$]{\includegraphics[width=4cm]{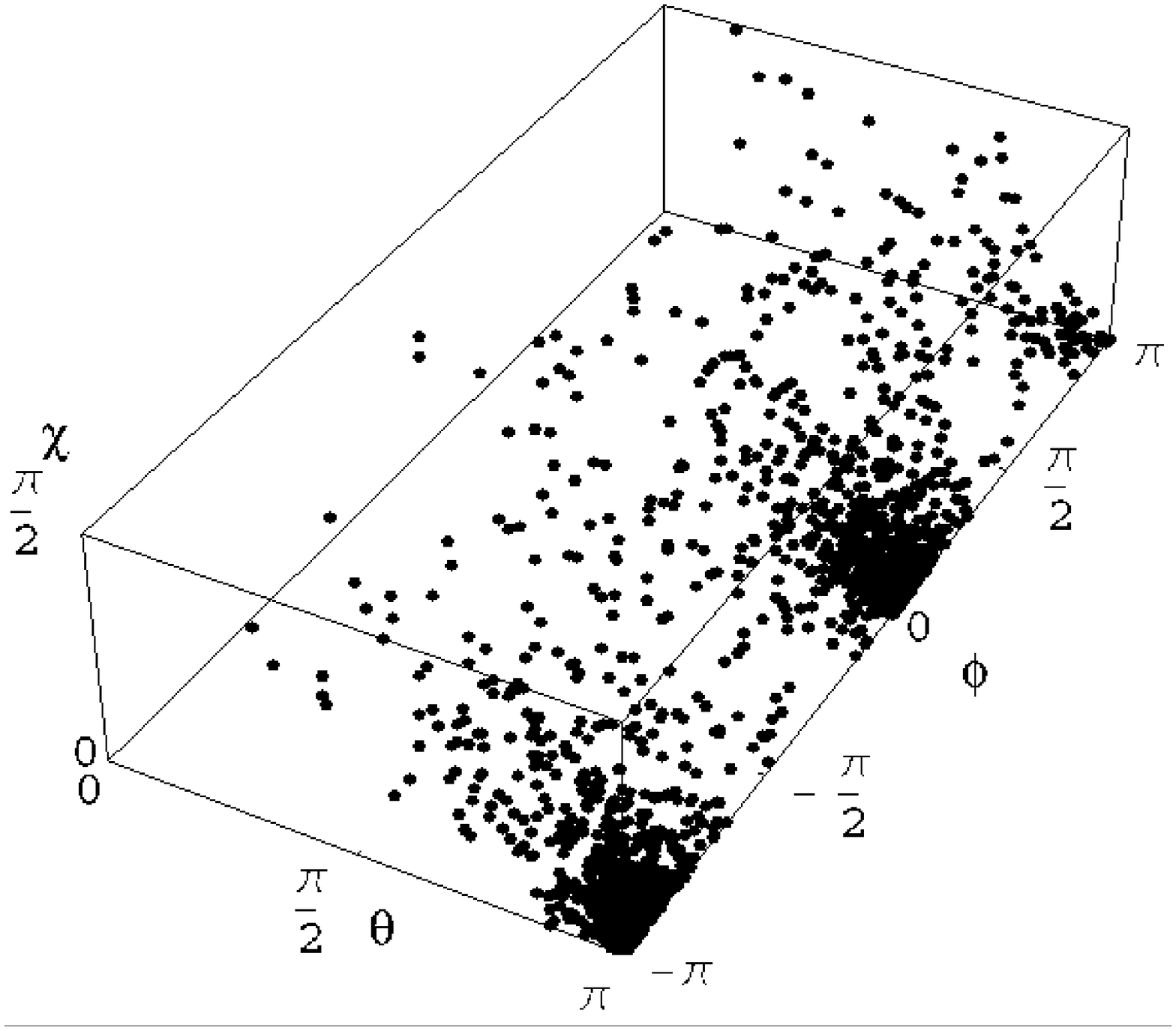}}
 \subfigure[$U^{(5)}$]{\includegraphics[width=4cm]{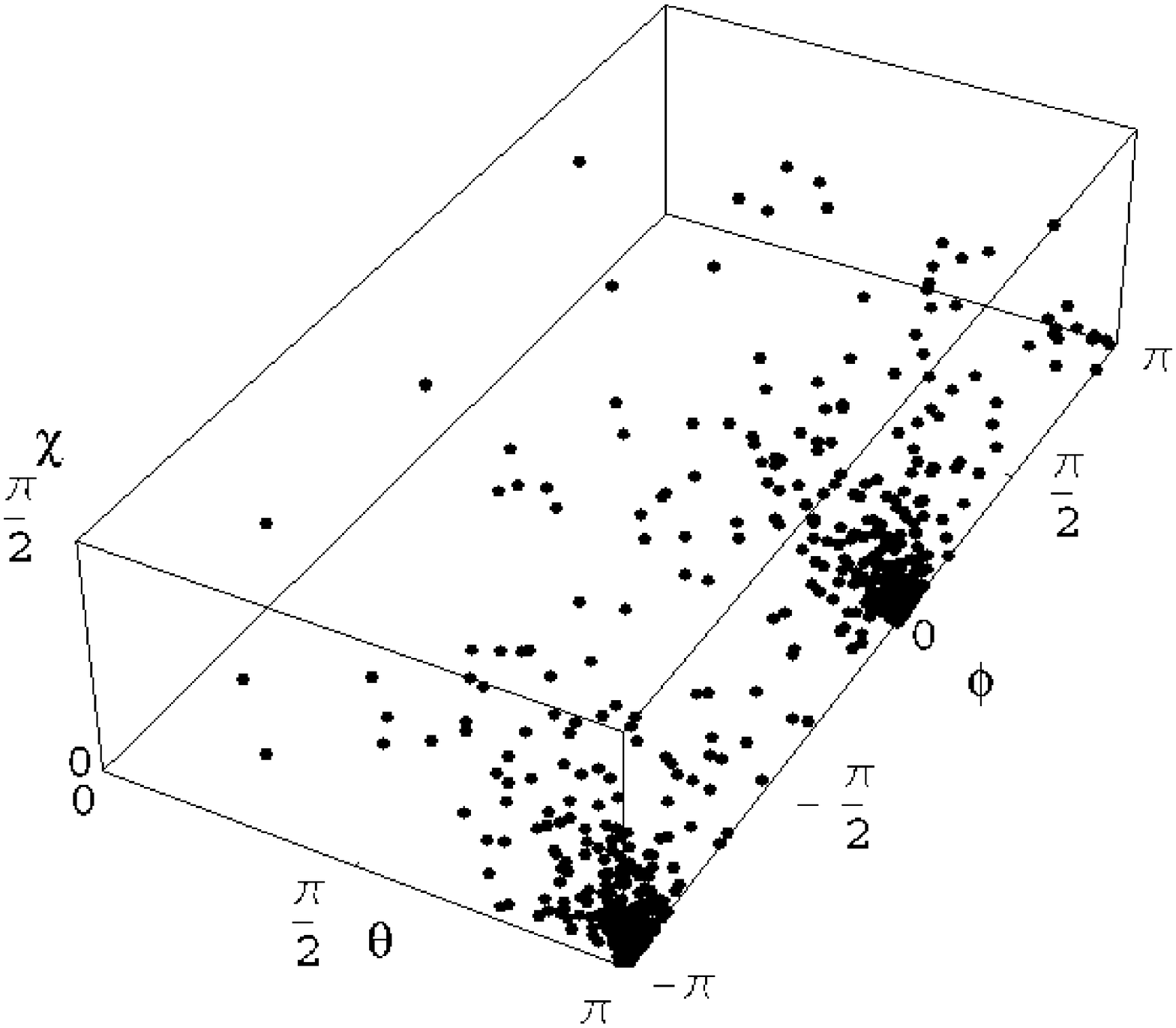}}
  \end{center}
\caption{(a) A set of rotations $U$ filling uniformly the space 
$(\theta,\phi,\chi)$ of parameters. Each point corresponds to a certain 
rotation $U$.  
(b)-(f) successive iterations of the mapping T(U) showing the convergence to 
the fixed point $U=-I$.
\label{f1}}
\end{figure}
Figure \ref{f1} shows the evolution of the parameters
$(\theta,\phi,\chi)$ upon successive transformations given by Eq. (\ref{T}). 
The set of initial rotations fills this space uniformly,
see Fig. \ref{f1} (a).
Figures \ref{f1} (b)-(f) show 
the successive iterations $U^{(1)}$ - $U^{(5)}$ of these points,
 illustrating the convergence of the mapping. 

From Eq. (\ref{it}) we can write down explicit formulas for the parameters 
$({\bf n}',\chi')$ of $U'=T(U)$ in terms of $({\bf n},\chi)$,

\[
\begin{split}
  \sin\chi' &= 4n_xn_y\sin^2\chi\sqrt{1-4n_x^2n_y^2\sin^4\chi}, \\
  n'_x &= -\frac{2n_y\sin\chi\cos\chi}{\sqrt{1-4n_x^2n_y^2\sin^4\chi}}, \\
  n'_y &= -\frac{2n_yn_z\sin^2\chi}{\sqrt{1-4n_x^2n_y^2\sin^4\chi}}, \\
  n'_z &= -\frac{1-2n_y^2\sin^2\chi}{\sqrt{1-4n_x^2n_y^2\sin^4\chi}}. \\
\end{split}
\]
We see that $\sin\chi=n_x=n_y=0$, $n_z=1$ is a fixed point, and this represents
 the identity operator. 
The stability of the fixed point can be analyzed by expanding close to the 
fixed point in the independent 
small quantities $\epsilon_s = \sin\chi/\sqrt{2}$, 
$\epsilon_x = n_x/\sqrt{2}$ and $\epsilon_y = n_y/2$.
We get 

\[
\begin{split}
 \epsilon_s' = \epsilon_x\epsilon_y\epsilon_s^2,\\
 \epsilon_x' = -\epsilon_y\epsilon_s,\\
 \epsilon_y' = -\epsilon_y\epsilon_s^2.
\end{split}
\]
were the primed quantities refer to the transformed rotation $U'$. 
It is clear that if $\epsilon_s,\epsilon_x,\epsilon_y$ are 
small quantities, then  $\epsilon_s',\epsilon_x',\epsilon_y'$ are even 
smaller and 
the fixed point is locally stable. 

\section{Convergence properties}

We studied numerically the convergence of the mapping $T(U)$ for points that 
are not close to the fixed point. 
Looking at Figure \ref{f1} we see that
although most points converge to the vicinity of the fixed point in a few 
iterations, there are some points that do not converge fast. To get a 
better understanding we do as follows. Let us choose some initial 
rotation angle $\chi$ and for each point in the $(\theta,\phi)$ plane 
construct the sequence $U^{(1)}$, $U^{(2)}$, ... stopping when 
$U^{(n)}$ is within a specified distance from the identity (we used 
the stopping criterion  $\chi^{(n)}<10^{-3}$). 
\begin{figure}[h]
  \begin{center}
 \subfigure[$\chi = 0.5$.]{\includegraphics[width=4cm]{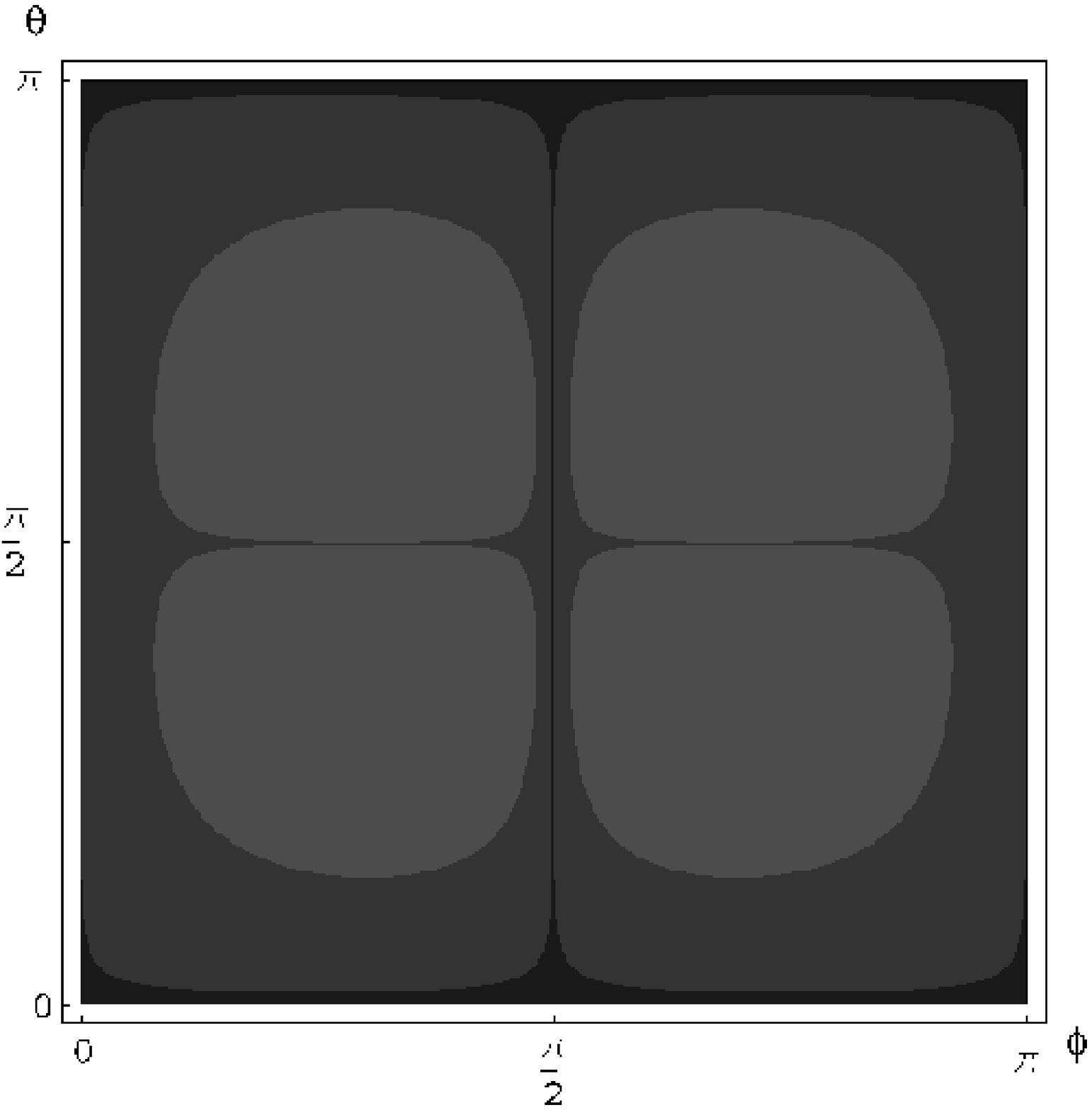}}
 \subfigure[$\chi = 1.0$.]{\includegraphics[width=4cm]{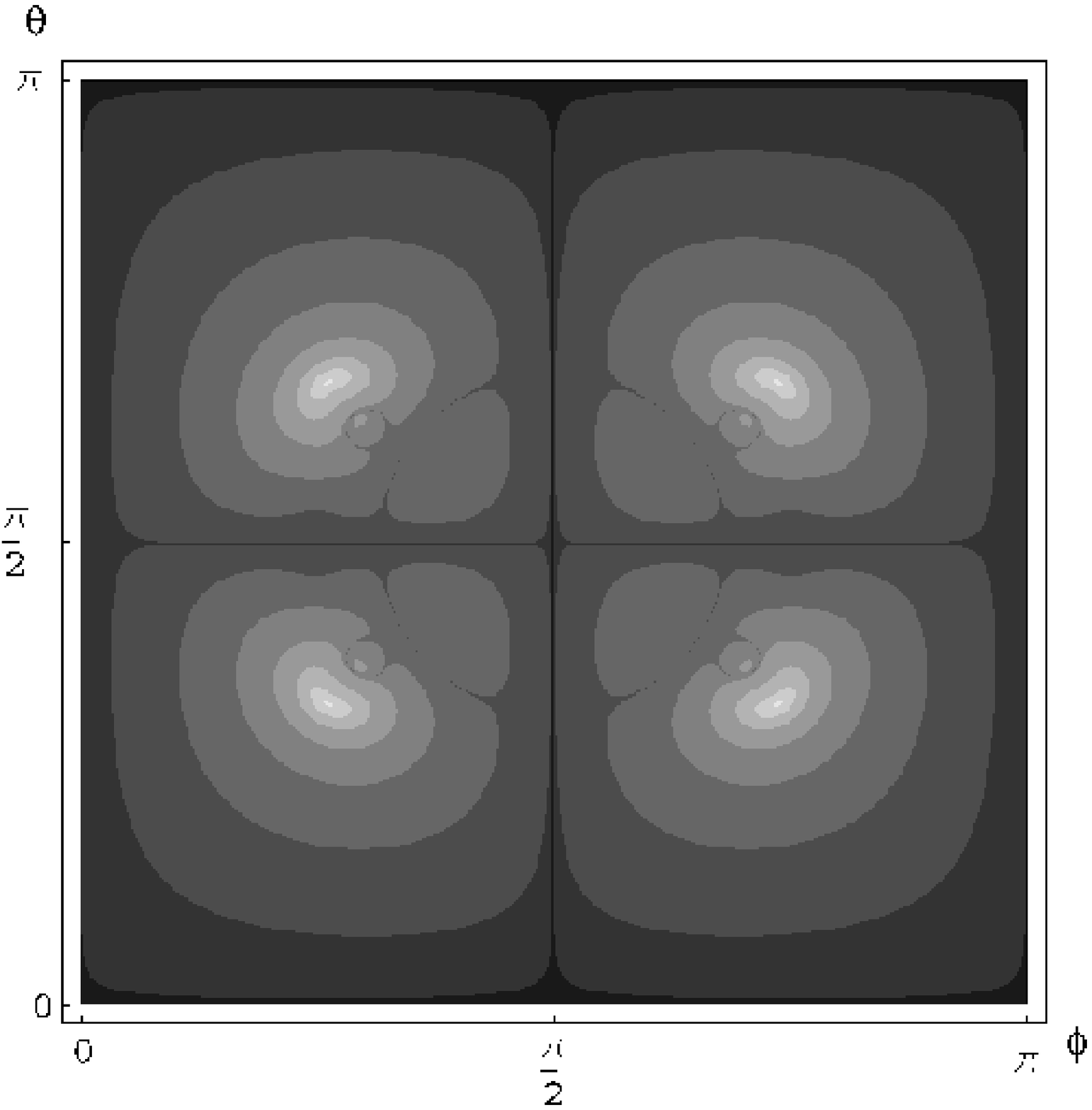}}\\
 \subfigure[$\chi = 1.25$.]{\includegraphics[width=4cm]{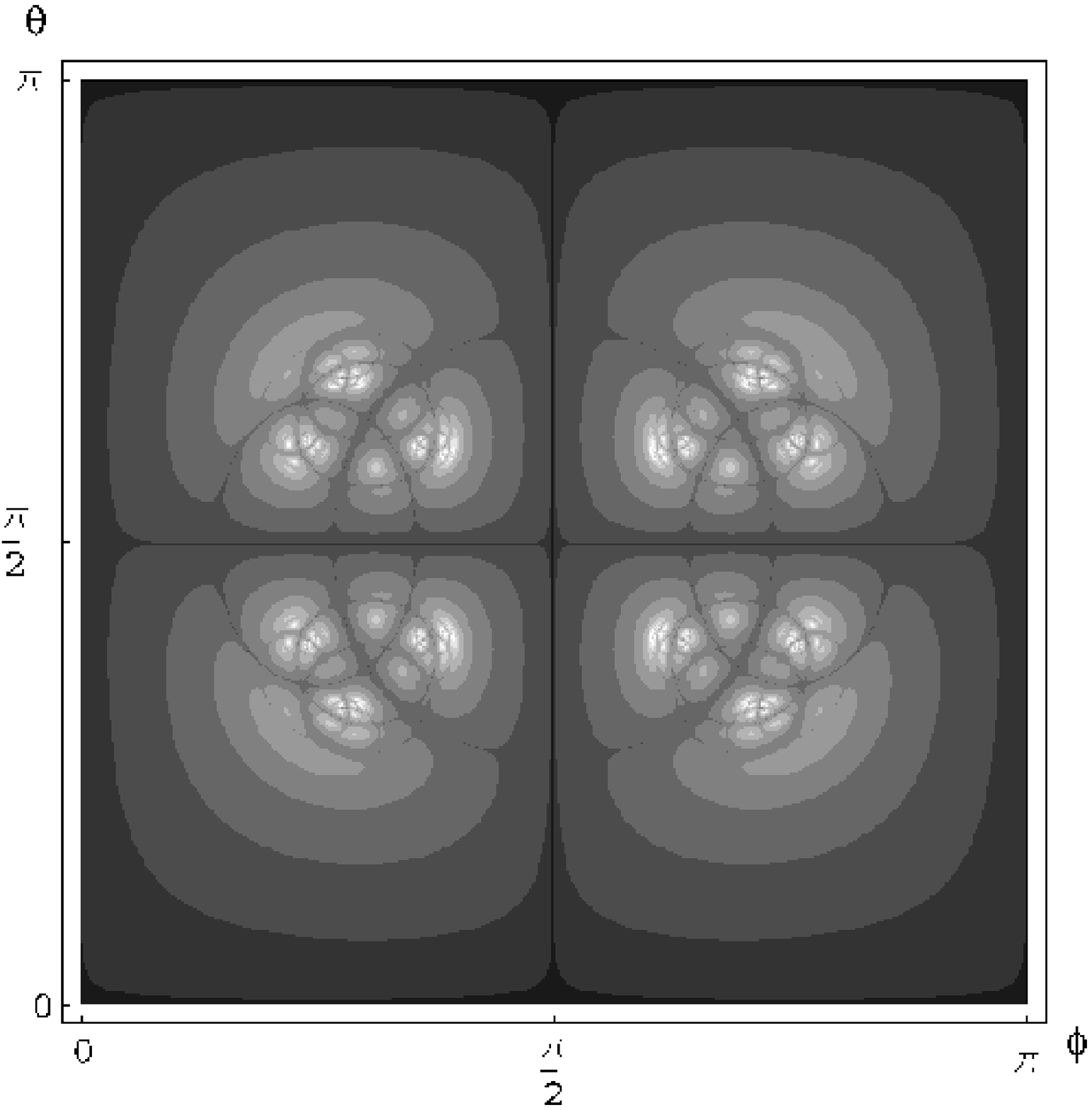}}
 \subfigure[$\chi = 1.5$.]{\includegraphics[width=4cm]{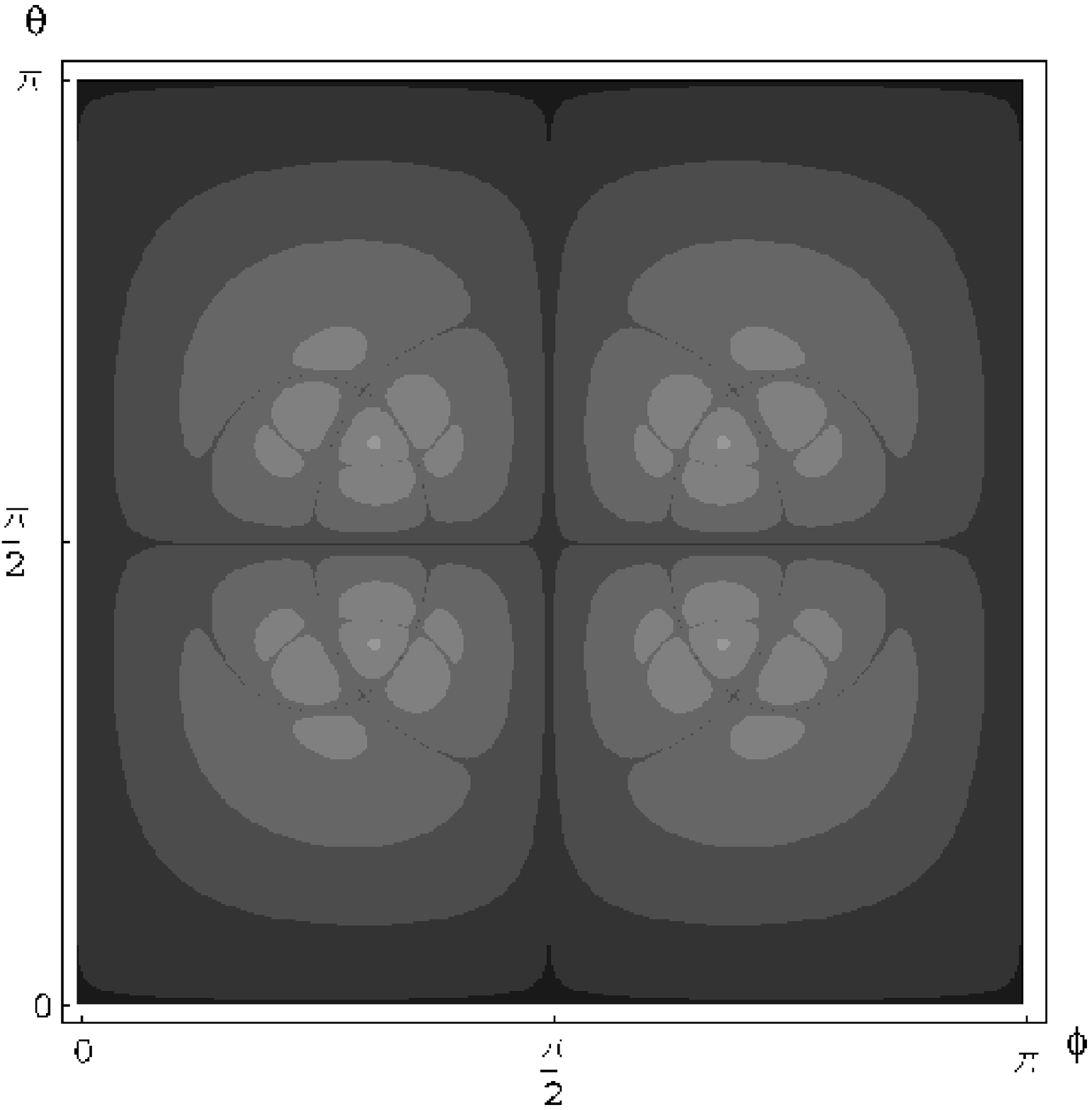}}\\
 \subfigure{\includegraphics[width=7cm]{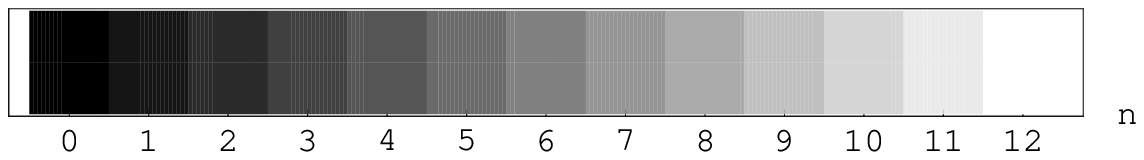}}\\
  \end{center}
\caption{The levels of gray indicate the number $n$  of iterations that
is needed to reach a rotation $U^{(n)}$ with the rotation 
angle $\chi^{(n)}<10^{-3}$. The darker shade means faster
convergence according to the scale at the bottom. 
\label{f2}}
\end{figure}
Figure \ref{f2} shows the $(\theta,\phi)$ plane for various 
$\chi$, the shades of gray representing the number $n$ of iterations 
needed for convergence. We see that for small $\chi$ convergence is fast 
and the pattern is simple, but for larger initial $\chi$ the pattern 
of convergence time 
is quite complex, and that there exist ``hard'' points, i.e. initial rotations
$({\bf n},\chi)$ that 
need a large number $n$ if iterations to converge.

\begin{figure}[h]
  \begin{center}
\includegraphics[width=8cm]{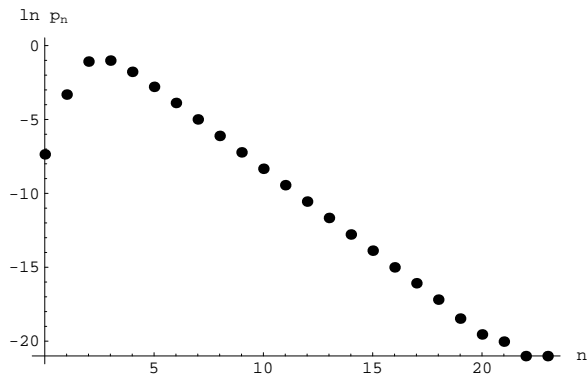}
  \end{center}
\caption{The fraction $p_n$ of the initial random set of rotations
  which did not reach the condition $\chi<10^{-3}$ till the $n$-th
  iteration.
\label{f5}} 
\end{figure}

To estimate the fraction of initial rotations that need a certain
number of iterations to converge to the fixed point we do the
following.  We start with an ensemble of random initial rotations $U$
characterized by a unit vector ${\bf n}$ distributed uniformly on a
sphere and by an angle $\chi$ distributed uniformly from in the inerval
$[0,\pi]$. We proceed with the iterations, Eq.~(\ref{T}), until
$\chi^{(n)}<10^{-3}$ is reached. The logarithm of the fraction
$p_n$ of initial rotations $U$ that needs $n$ iterations to converge is shown
in Figure \ref{f5} as a function of n.
We find that about 74\% of the initial rotations will converge after
applying the third iterate $U^{(3)}$ and about 91\% do so after the iterate 
$U^{(4)}$ .
Observe that except for the first points, all points fall on a
straight line which means that the fraction $p_n$ of ``difficult''
initial rotations decays exponentially with the number of iterations,
\begin{equation}\label{a}
  p_n \propto e^{-\alpha n}, \qquad \alpha\approx 1.1.
\end{equation}
If $k$ is the number of repetitions of $U$ in $T(U)$ (for the mapping
(\ref{T}) we have $k=4$ ), the total time of the sequence $U^{(n)}$ is
$t=\tau k^n$, where $\tau$ is the time between the pulses. The relation
(\ref{a}) can then also be written as the fraction $p(t)$ of the initial
rotations which did not converge till time $t$,
\begin{equation}
\label{t}
p(t)\propto\left(\frac{t}{\tau}\right)^{-\beta},\qquad
 \beta=\frac{\alpha}{\ln k}.
\end{equation}
The relations Eqs.~(\ref{a}) and (\ref{t}) are established by running
a simulation with the use of a specific convergence criterion, $\chi^{(n)}
<10^{-3}$. Changing the criterion affects the proportionality
coefficients in these relations, but does not change the values of
exponents $\alpha$ and $\beta$.

\begin{figure}[h]
  \begin{center}
\includegraphics[width=8cm]{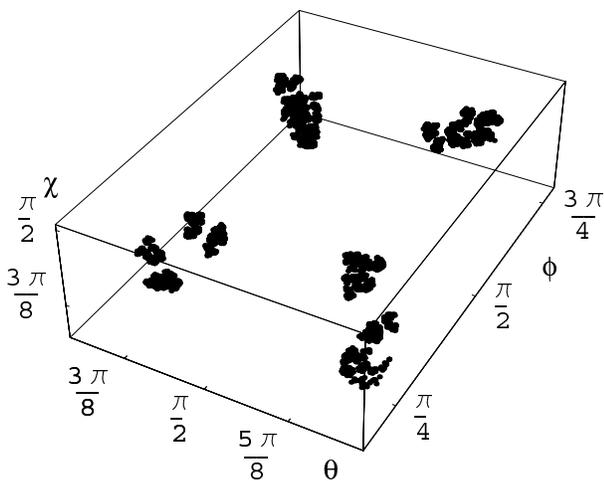}
  \end{center}
\caption{The set $M_9$ of initial rotations that do not
  converge in nine iterations, according to the criterion $\chi <10^{-3}$.
\label{f3}} 
\end{figure}

Let $M_n$ denote the set of initial rotations that do not converge in 
$n$ iterations. 
[As an example, the domain $M_9$ of ``difficult''  initial rotations
(according to the criterion $\chi <10^{-3}$) 
is presented in Fig.~\ref{f3}]. 
The sequence of sets $M_n$ is such that each set is contained in the 
previuos one, $M_n\subset M_{n-1}$, and the fraction of points $p_n$ in Figure 
\ref{f5} is proportional to the difference in the volumes of $M_{n+1}$ and 
$M_{n}$. The set $M=\bigcap_{n=0}^\infty M_n$ 
of ``infinitely hard'' points
appears to be a fractal with fractal dimension $D\approx1.5$ 
according to the box-counting algorithm we used.

\section{Discussion}

We considered the possibility of constructing an echo pulse sequence that does
not require applying a high permanent field to the system. The advantage 
of such a method is that between the pulses the dynamics of the system is not 
influenced by external perturbations. In the context of ESR of a quantum dot, 
this method may help distinguishing between different effects of the 
hyperfine interaction: this interaction creates some random effective 
magnetic field acting on the electron spin, but may also lead to electron
spin relaxation. The absence of the external permanent field in this problem 
is crucial, as its application definitely suppresses the spin relaxation 
part of the hyperfine interaction effects \cite{Khaetskii}. 

We have provided a solution to the general echo problem in terms of a
set of longer and longer pulse sequences that give successively
better approximations to the identity operator. We have tested several longer
sequences {\it e.g} $T(U) = ZUXUYUXUZUXUYUXU$.
In all cases we
found that the exponent $\beta$ in Eq.~(\ref{t}) is independent of the
particular mapping chosen.  Whether this represents some inherent
property of the problem or only is the case for the limited class of
mappings we have studied is not known. For example, we have not
studied sequences with control pulses other than $\pi$ rotations about
the coordinate axes.
We have also not ruled out the
possibility of a  solution that will yield an ideal echo with the help of 
a finite number of control pulses.

The idea of using iterated mappings to generate pulse sequences has been 
used in NMR applications \cite{Tycko} but as far as we know this 
particular problem or the mapping we study was never discussed. The 
mapping we have used was also proposed in Ref. \onlinecite{Lidar2005} 
in the context of dynamical decoupling of a qubit.

\acknowledgments This work was supported by the Norwegian Research
Council via a StorForsk program and by NSF grants DMR 02-37296 and DMR
04-39026 at the University of Minnesota. The hospitality of
MPIPKS-Dresden (L.I.G.) is gratefully acknowledged.

\end{document}